\documentclass[a4paper,twocolumn,english,8pt,amssymb,nofootinbib,superscriptaddress]{revtex4}
\usepackage{lmodern}
\usepackage{lmodern}

\usepackage[T1]{fontenc}
\usepackage[latin9]{inputenc}
\usepackage{babel}
\usepackage{mathrsfs}
\usepackage{amsmath}
\usepackage{amssymb}
\usepackage{esint}
\usepackage[unicode=true,pdfusetitle,
 bookmarks=true,bookmarksnumbered=false,bookmarksopen=false,
 breaklinks=false,pdfborder={0 0 1},backref=false,colorlinks=false]
 {hyperref}

\makeatletter

\@ifundefined{textcolor}{}
{%
 \definecolor{BLACK}{gray}{0}
 \definecolor{WHITE}{gray}{1}
 \definecolor{RED}{rgb}{1,0,0}
 \definecolor{GREEN}{rgb}{0,1,0}
 \definecolor{BLUE}{rgb}{0,0,1}
 \definecolor{CYAN}{cmyk}{1,0,0,0}
 \definecolor{MAGENTA}{cmyk}{0,1,0,0}
 \definecolor{YELLOW}{cmyk}{0,0,1,0}
}

\usepackage{babel}
\usepackage{babel}
\usepackage{babel}
\usepackage{babel}
\usepackage{babel}
\usepackage{babel}
\usepackage{babel}
\usepackage{babel}
\usepackage{babel}
\usepackage{babel}
\usepackage{babel}
\usepackage{babel}
\usepackage{babel}
\usepackage{babel}
\usepackage{babel}
\usepackage{babel}

\usepackage{babel}
\usepackage{graphicx}
\def\b{\begin{equation}}
\def\e{\end{equation}}

\@ifundefined{textcolor}{}{%
 \definecolor{BLACK}{gray}{0}
 \definecolor{WHITE}{gray}{1}
 \definecolor{RED}{rgb}{1,0,0}
 \definecolor{GREEN}{rgb}{0,1,0}
 \definecolor{BLUE}{rgb}{0,0,1}
 \definecolor{CYAN}{cmyk}{1,0,0,0}
 \definecolor{MAGENTA}{cmyk}{0,1,0,0}
 \definecolor{YELLOW}{cmyk}{0,0,1,0}
 }

\usepackage{latexsym}\usepackage{bm}

\makeatother

\begin{document}
\title{ Hyperbolic Metamaterials and Massive Klein-Gordon Equation in 2+1 Dimensional de Sitter Spacetime}
\author{Bayram Tekin}
\email{btekin@metu.edu.tr}
\affiliation{Department of Physics,\\
 Middle East Technical University, 06800 Ankara, Turkey}
\selectlanguage{english}%

\begin{abstract}
\noindent The wave equation obeyed by the extraordinary component of the electric field in a hyperbolic metamaterial was shown to be a {\it massless } Klein-Gordon field living in a flat spacetime with two timelike and two spacelike dimensions. Such a wave equation, unexpectedly, allows dispersionless propagation albeit having two spatial dimensions. Here we show that the same equation can be naturally interpreted as a particular  {\it massive} Klein-Gordon equation with  the usual one timelike and two spacelike dimensions in a de Sitter (dS) background spacetime.  The mass parameter of the scalar field is given in terms of the cosmological constant, Planck constant and the speed of light as $m =\sqrt{\Lambda}\frac{\hbar}{c}$ which corresponds to the point for which the left and right conformal weights of the boundary conformal field theory (CFT) (via the de Sitter/CFT correspondence) are equal. This particular mass corresponds to the gapless mode in the dS spacetime for which the dispersion relation is linear in the wave number.
\end{abstract}
\maketitle
\section{Introduction}

 In a $2+1$ dimensional flat world, one cannot send wave pulses without dispersion in a vacuum. That means any wave pulse will broaden and change shape while propagating even if all frequencies move at the same wave speed. Thus two agents in communication  will probably misconstrue their messages unless they are well-versed in the solutions of the wave equation in 2 spatial dimensions. 

This is in sharp contrast to the 3+1 dimensional world we are living in: in a vacuum, a light pulse does not disperse, while in a material medium with a frequency-dependent  refraction index, it does. In fact, in all odd spatial dimensions (except one dimension), the wave equation yields dispersionless propagation in a vacuum, while for all even spatial dimensions it yields a dispersive vacuum.  One spatial dimension is a rather exceptional case: one can design initial data for which there is no dispersion, but generically 1+1 dimensional wave equation is dispersive. This dimension-dependent dispersion is called {\it anomalous dispersion} \cite{CH}.

This was the state of affairs until recently, in a remarkable work by Bender {\it et al.} \cite{Bender}, it was shown that while the 2+1 dimensional vacuum is dispersive,  2+2 dimensional vacuum need not be so! Thus, for the first time, these authors proved that adding  one more {\it timelike} direction can solve the problem of anomalous dispersion and eliminate the violation of the Huygens' principle. The way they have shown this was, as we shall describe below, via the introduction of a modified wave equation with non-constant (time-dependent) coefficients which upon transformation gives the flat space wave equation in a spacetime with 2+2 dimensions.  
Of course, one might think that such a solution would be a purely mathematical construction since an extra time dimension is not available for purchase; but  hyperbolic metamaterials are. It turns out that an additional effective timelike dimension appears naturally in a metamaterial with a hyperbolic dispersion which was introduced in the ground-breaking work \cite{SN}.  Smolyaninov and Narimanov showed that in a nondispersive, nonmagnetic, uniaxial anisotropic metamaterial the "extraordinary" component of the electric field obeys a {\it massless } Klein-Gordon equation in a flat 2+2 dimensional spacetime. Considering the works \cite{SN,Bender} together opens a new window of dispersionless communication in flatland: all one needs to do is to consider a hyperbolic metamaterial flatland which effectively is a 2+2 dimensional world.  

In this work, we will show that another way to interpret the wave equation in a hyperbolic metamaterial is to consider a background {\it de Sitter } spacetime with the usual 1 timelike and 2 spacelike directions; and in this curved background spacetime, the extraordinary component of the electric field obeys a {\it massive} Klein-Gordon equation with the usual $(-,+,+)$ signature. But the mass parameter of the scalar field must be tuned to the cosmological constant of the background spacetime. This particular tuned mass yields a gapless mode for which the dispersion relation is linear in the wave number.
Hence we propose another way to circumvent the anomalous dispersion problem in 2 spatial dimensions and restore the Huygens' principle. This particular 
curved background interpretation of the hyperbolic metamaterial can yield interesting connections along the line of dS/CFT correspondence \cite{Strominger}. The mass turns out to yield equal left and right conformal weights in the boundary conformal field theory.

The layout of the paper is as follows: In Section II, we briefly recapitulate the relevant 2+2 dimensional wave equation in flat spacetime obeyed by the extraordinary component of the electric field in a metamaterial; in Section III, we discuss the mapping  of the modified wave equation with non-constant coefficients to the 2+2 dimensional flat space equation. In Section IV, we show that the mentioned modified wave equation, which was introduced by guess in an {\it ad hoc}  manner actually comes from a scalar field theory in de Sitter  spacetime with a tuned mass parameter. In that section we also show how the surprising dispersionless propagation is achieved in 2 spatial dimensions. 

\section{Extraordinary component of the electric field in a metamaterial as a massless Scalar Field in $2+2$ dimensions}

In \cite{SN}, it was shown that, in a material with properties as described in the Introduction section above, and 
with a dielectric permitivity tensor ${\bf{\varepsilon}} = \text{diag}({\varepsilon_x,\varepsilon_x, \varepsilon_z} )$, the extraordinary component of the electric field satisfies the massless Klein-Gordon equation 
\begin{equation}
\frac{1}{c^2}\frac{\partial^2 \varphi}{\partial t^2} = \frac{1}{\varepsilon_x}\frac{\partial^2 \varphi}{\partial z^2}+ \frac{1}{\varepsilon_z}\left(\frac{\partial^2 \varphi}{\partial x^2}+\frac{\partial^2 \varphi}{\partial y^2} \right).
\label{KG1}
\end{equation}
For $\varepsilon_x <0$ and $\varepsilon_z >0$, one has an indefinite metamaterial with interesting optical properties. See \cite{Sm2} for a nice review. Clearly in this case the massless Klein-Gordon equation (\ref{KG1}) lives in a flat four dimensional space with the signature  $(-,-,+,+)$. Spacetimes with two timelike directions appear in some higher dimensional theories \cite{Bars}. Having only two spatial dimensions at our disposal, the question arises if light pulses can propagate without dispersion in this space. 

\section{Modified wave equation with nonconstant coefficients}

The lore for linear wave equations in a spacetime with even number of spatial dimensions and a single time dimension is that they have anomalous dispersion even though all the frequencies propagate with the same speed and hence all chromatic dispersion is eliminated. This obstruction to keeping a pulse's shape intact would be somewhat a disadvantage for the use of effectively 2D materials. But it was shown that  equation (\ref{KG1}) is equivalent to a  {\it modified} wave equation as given by equation (3) in \cite{Bender} which is of the form 
\begin{equation}
\frac{\partial^2 u}{\partial \tau^2} -\frac{1}{\tau} \frac{\partial u}{\partial \tau}+\frac{1}{\tau^2}u = c^2 \left(\frac{\partial^2 u}{\partial X^2}+\frac{\partial^2 u}{\partial Y^2} \right).
\label{KG2}
\end{equation}
Note that, in contrast to the discussion in \cite{Bender}, we are using different coordinates in (\ref{KG1}) and  (\ref{KG2}) to make the correspondence clear.
Equivalence of equations (\ref{KG1}) and (\ref{KG2}) follows in three steps: with the identifications  
$ u (\tau,X,Y) :=  \tau v(\tau,X,Y)$,  (\ref{KG2}) becomes
\begin{equation}
\frac{\partial^2 v}{\partial \tau^2} +\frac{1}{\tau} \frac{\partial v}{\partial \tau} = c^2 \left(\frac{\partial^2 v}{\partial X^2}+\frac{\partial^2 v}{\partial Y^2} \right).
\label{KG3}
\end{equation}
Then defining a two dimensional vector time variable $\vec{\tau} := (\alpha, \beta)$, the left-hand side of the last equation is the radial part of the Laplacian in 2D, hence in the  Cartesian $(\alpha, \beta)$ coordinates, (\ref{KG3}) becomes
\begin{equation}
\frac{1}{c^2}\left(\frac{\partial^2 v}{\partial \alpha^2} + \frac{\partial^2 v}{\partial \beta^2}\right)=  \frac{\partial^2 v}{\partial X^2}+\frac{\partial^2 v}{\partial Y^2}.
\label{KG4}
\end{equation}
To match (\ref{KG1}) and (\ref{KG4}), we set  $\alpha = \tau$, $\beta = \sqrt{|\epsilon_x|} \frac{ z}{c}$, $X = \sqrt{\epsilon_z} x$, $Y = \sqrt{\epsilon_z} y$ and $v = \varphi$. With these identifications and considering an initial non-zero disturbance having a vanishing derivative as the other initial condition, one has a dispersionless propagation \cite{Bender}. It will be clear why this equation yields dispersionless propagation in the next section.

\section{Scalar Field in de Sitter Spacetime with a tuned mass parameter}

Let us now show that the modified wave equation (\ref{KG2}) which was introduced by guesswork and hence its two time two space massless Klein-Gordon version (\ref{KG1}) are equivalent to a minimally coupled "massive" Klein-Gordon equation with a tuned mass in a 2+1 dimensional de Sitter background.  Consider the de Sitter metric in Poincar\'{e} coordinates
\begin{eqnarray}
ds^{2}=\frac{\ell^{2}}{c^2 \tau^{2}}\left(-c^2 d \tau^{2}+dX^{2}+dY^{2}\right),
\label{met}
\end{eqnarray}
where $\ell$  being the dS radius related to the positive cosmological constant as $ \Lambda = \frac{1}{\ell^{2}}$. It is a maximally symmetric space with the  Ricci tensor given as $R_{\mu \nu} = 2 \Lambda g_{\mu \nu}$. [In 3 dimensions the Riemann and the Ricci tensor carry the same amount of information, hence the full curvature is determined by the Ricci tensor.]
In this background, consider a minimally coupled massive scalar field \cite{Tekin}
\begin{eqnarray}
I&=&-\frac{1}{2}\int d^{3}X\,\sqrt{-g}\left(\partial_{\mu}\Phi\partial^{\mu}\Phi+\left(\frac{m c}{\hbar}\right)^2\Phi^{2}\right)\label{action} \\
&=&-\frac{1}{2}\int d^{3}X\,\left\{ \frac{\ell}{c \tau}\left[-\frac{1}{c^2}\dot{\Phi}^{2}+\left(\partial_{i}\Phi\right)^{2}\right]+\frac{\ell^{3}}{c \hbar^2\tau^{3}}m^{2}\Phi^{2}\right\},  \nonumber
\end{eqnarray}
where in the second line we inserted the metric (\ref{met}) and defined $\dot{\Phi} \equiv \frac{\partial \Phi}{\partial \tau}$; $\partial_{i}\Phi = (\partial_{X}\Phi,\partial_{Y}\Phi )$. Then variation of the action (\ref{action}) with respect to $\Phi$ yields 
\begin{equation}
\frac{\ell}{c \tau}\left(-\frac{1}{c^2}\ddot{\Phi}+\frac{1}{c^2 \tau}\dot{\Phi}+\partial^{2}\Phi\right)-\frac{\ell^{3}}{c \hbar^2 \tau^{3}}m^{2}\Phi=0,
\label{KG5}
\end{equation}
which is just the massive Klein-Gordon equation written covariantly as $ \left(\Box-\left(\frac{m c}{\hbar}\right)^2\right)\Phi=0$ where $\Box$ is the d'Alembert operator in the dS background; and $\partial^{2}= \frac{\partial^2 }{\partial X^2}+\frac{\partial^2 }{\partial Y^2}$. The first order of business is to obtain the modified wave equation (\ref{KG2}) from (\ref{KG5}), for this purpose we must identify the mass parameter of the Klein-Gordon field as
\begin{equation}
m = \frac{\hbar}{\ell c}=\sqrt{\Lambda}\frac{\hbar}{c} ,
\label{tuned_mass}
\end{equation}
and set $ \Phi = u$. So, rather remarkably, the 2+1 dimensional modified wave equation (\ref{KG2}) introduced in \cite{Bender} is related to apparently different wave equations : one is the massless Klein-Gordon equation (\ref{KG1}) that describes the extraordinary component of the electric field in a flat spacetime with $(-,-,+,+)$ signature; and the other is the massive Klein-Gordon equation with a tuned mass  in 2+1 dimensional de Sitter spacetime. These equations circumvent the anomalous dispersion problem that inflicts all spacetimes with an even number of spatial dimension. 

Let us explore further what special feature arises for the tuned mass (\ref{tuned_mass}) in de Sitter space. For this purpose, we consider Fourier mode type solutions, from which a wave packet can be constructed. To simplify the resulting equation, we define a new time coordinate $t$ by setting $\tau :=  \frac{\ell}{c} e^{ -\frac{c t}{\ell}}$. Then the de Sitter metric takes the form 
\begin{eqnarray}
ds^{2}=-c^2 d t^{2}+a(t)^2\left(dX^{2}+dY^{2}\right),
\label{met}
\end{eqnarray}
with $a(t):=   e^{ \frac{c t}{\ell}} $. Then  (\ref{KG5}) becomes
\begin{equation}
-\ddot{\Phi}-2\frac{\dot{a}}{a}\dot{\Phi}+\frac{c^2}{a^2}\partial^{2}\Phi
-\frac{c^4 m^2}{\hbar^2}\Phi=0.
\label{reduced}
\end{equation}
Consider one single Fourier mode which can be taken as (we shall deal with the reality of the scalar field in a moment) : 
\begin{equation}
\Phi(t, X,Y) := \frac{f_{\vec{k}}(t)}{a(t)}e^{i \vec{k}\cdot \vec{X}},
\end{equation} 
with $\vec{X} := (X,Y)$. Inserting this into (\ref{reduced}), one arrives at a harmonic oscillator equation
\begin{equation}
\ddot{f}_{\vec{k}}+ \omega^2_{\vec{k}} f_{\vec{k}}=0,
\end{equation}
where the dispersion relation reads
\begin{equation}
\omega^2_{\vec{k}} = -\frac{c^2}{\ell^2} + \frac{c^4 m^2}{\hbar^2}+ \frac{k^2 c^2}{a^2}.
\label{massbabam}
\end{equation}
So generically the group velocity $v^i_{\text{g}} = \frac{\partial  \omega_{\vec{k}}}{\partial k_i}$ depends on $k$ and every mode moves with a different speed leading to dispersion; this includes the $m=0$ modes.  The only exception is the the modes with  the tuned mass $m = \frac{\hbar}{\ell c}$, for which all the modes are gapless $\omega_{\vec{k}} =   \frac{k c}{a}$ as in the case of a massless particle in flat spacetime. The group velocity and the phase velocity are equal to each other and they are independent of $\vec{k}$. So the tuned mass parameter in de Sitter background is akin to the massless mode in flat spacetime.  A wave packet, pulse will not lose its shape when propagating from one point to another. For completeness let us write the generic solution to (\ref{reduced}) as
\begin{equation}
\Phi(t, X,Y)  = \int \frac{d^2\vec{k}}{ 2\pi a} \Bigg( c_{\vec{k}} f_{\vec{k}}(t) e^{i \vec{k}\cdot \vec{X}}+c^*_{\vec{k}} f_{\vec{k}}(t) e^{-i \vec{k}\cdot \vec{X}} \Bigg),
\end{equation}
where we have taken care of the reality of the scalar field;  $c_{\vec{k}}$ are arbitrary complex constants.

Let us note that (\ref{KG5}) was studied in detail in \cite{Strominger} in planar coordinates in the context of defining a dual conformal field theory in the future boundary of bulk $dS_3$ with a massive scalar field. For a unitary conformal field theory in this two dimensional boundary, the bulk scalar field corresponds to boundary operators with real weights which is possible only if  the mass of the scalar field satisfies  $0 < m < \frac{\hbar}{\ell c}$. The conformal weights of the 2D Euclidean unitary boundary conformal field theory are given as $h_{\pm} = 1\pm \sqrt{ 1-  \frac{c^2\ell^2 m^2}{\hbar^2 }}$. We have now seen that if the tuned mass (\ref{tuned_mass}) remains outside the unitarity region and corresponds to the degenerate case of $h_+ = h_-$. The crucial point is that this particular massive Klein-Gordon equation in the bulk describes a hyperbolic metamaterial in which there is no anomalous dispersion.

\section{Conclusions}

We have shown that the extraordinary component of the electric field in a metamaterial which is known to obey a wave equation a  flat 2+2 dimensional space; and which was shown to be equivalent to a modified wave equation with time-dependent coefficients, naturally arises as a particular massive scalar field in a 2+1 dimensional de Sitter spacetime. The mass parameter of the scalar field yields gapless modes and the dispersion relation between the angular frequency and the wave number is linear resulting a dispersionless propagation. It is important to realize that this happens in de Sitter spacetime and not in anti-de Sitter spacetime as the minus sign in the first term of (\ref{massbabam}) is important. It is likely that this metamaterial presents and example of the dS/CFT conjecture but  identification of the two dimensional conformal field theory corresponding to the hyperbolic metamaterial in this context is an outstanding  problem. 

{\it Acknowledgments.} We would like to thank  Z. T. Ozkarsligil, T.C. Sisman and A. Karasu for useful discussions.

\end{document}